\documentclass[12pt]{article}

\usepackage{scicite}
\usepackage{soul}
\usepackage{xcolor}
\usepackage{amsmath}

\usepackage{times}

\newcommand{\Yb}{$^{171}$Yb}

\newcommand{\ket}[1]{\left|  #1 \right\rangle}
\newcommand{\aver}[1]{\ensuremath{\langle {#1} \rangle}}

\usepackage{graphicx}

\newenvironment{sciabstract}{%
\begin{quote} \bf}
{\end{quote}}

\title{Time-Reversal-Based Quantum Metrology with Many-Body Entangled States}

\author
{Simone Colombo,$^{1\ast}$ Edwin Pedrozo-Pe\~{n}afiel,$^{1\ast}$ Albert F. Adiyatullin,$^{1\ast}$ \\ Zeyang Li,$^{1}$ Enrique Mendez,$^{1}$ Chi Shu,$^{1,2}$ Vladan Vuleti\'{c}$^{1}$\\
\\
\normalsize{$^{1}$Department of Physics, MIT-Harvard Center for Ultracold Atoms}\\ 
\normalsize{and Research Laboratory of Electronics,} \\
\normalsize{Massachusetts Institute of Technology,}\\
\normalsize{$^{2}$Department of Physics, Harvard University}\\
\normalsize{Cambridge, Massachusetts 02139, USA}\\
\\
\normalsize{$^\ast$These authors contributed equally to this work}
}

\date{}

\begin{document} 

\setstcolor{red}

\baselineskip24pt


\maketitle


\begin{sciabstract}

In quantum metrology, entanglement represents a valuable resource that can be used to overcome the Standard Quantum Limit (SQL) that bounds the precision of sensors that operate with independent particles. Measurements beyond the SQL are typically enabled by relatively simple entangled states (squeezed states with Gaussian probability distributions), where quantum noise is redistributed between different quadratures. However, due to both fundamental limitations and the finite measurement resolution achieved in practice, sensors based on squeezed states typically operate far from the true fundamental limit of quantum metrology, the Heisenberg Limit. Here, by implementing an effective time-reversal protocol through a controlled sign change in an optically engineered many-body spin Hamiltonian, we demonstrate atomic-sensor performance with non-Gaussian states beyond the limitations of spin squeezing, and without the requirement of extreme measurement resolution. Using a system of 350 neutral {\Yb} atoms, this signal amplification through time-reversed interaction (SATIN) protocol achieves the largest sensitivity improvement beyond the SQL ($11.8 \pm 0.5$~dB) demonstrated in any (full Ramsey) interferometer to date. Furthermore, we demonstrate a precision improving in proportion to the particle number (Heisenberg scaling), at fixed distance of 12.6~dB from the Heisenberg Limit. These results pave the way for quantum metrology using complex entangled states, with potential broad impact in science and technology. Possible future applications include searches for dark matter and for physics beyond the standard model, tests of the fundamental laws of physics, timekeeping, and geodesy.

\end{sciabstract}


Over the last two decades, substantial effort has been devoted towards the design of protocols and the engineering of quantum states that enable the operation of atomic sensors beyond the Standard Quantum Limit (SQL) \cite{Takano2009,Appel2009,Sewell2012,Hamley2012,berrada2013integrated,muessel2014Scalable,schmied2016bell,Cox2016a,Hosten2016,Hosten2016a,Bohnet2016,braverman2019near,Bao2020,Pedrozo2020entanglement,huang2020self,kruse2016improvement}.
The SQL arises from the discreteness of outcomes in the quantum measurement process, i.e. the quantum projection noise, and sets the limit of precision $1/\sqrt{N}$ that can be achieved with a system of $N$ independent particles. The  SQL can be overcome by generating many-body entanglement, most commonly achieved by means of spin squeezing \cite{Kitagawa1993, Wineland1994}, where a state of the collective spin with reduced quantum noise along one quadrature is created and detected. Such an approach is often limited by the precision of the readout rather than the generation of the squeezed state \cite{braverman2019near,Cox2016a,Hosten2016,Appel2009}.

The ultimate boundary for linear quantum measurements is the Heisenberg Limit (HL), where the precision improves with particle number as $1/N$. The HL can be reached with maximally entangled states, or equivalently, when the quantum Fisher information $F$ of the system is the largest~\cite{Pezze2018}. Maximally entangled states have been generated, but only in relatively small systems of up to 20 particles \cite{Leibfried2005,Monz-Blatt2011, omran2019CatState,DiCarlo-CatSupercond,Song-10CatSuperCond} and at reduced fidelity, and they are extremely difficult to create and maintain in many-atom systems that are of interest for metrological applications. As an alternative, more easily implementable schemes and quantum states have been identified where the precision improves as $b/N$ (Heisenberg scaling, HS \cite{Saffman2009,Davis2016, Frowis2016}), at fixed distance $b \geq 1$ from the HL. One such approach is to create an entangled state with large quantum Fisher information via a Hamiltonian process, then subject the system to the signal to be measured (i.e. a phase shift $\varphi$) before evolving it ''backwards in time'' by applying the negative Hamiltonian. This Loschmidt-echo-like approach \cite{Toscano2006,Davis2016,Frowis2016,Nolan2017,Macri2016} results in a final state that is displaced relative to the initial state, and where under appropriate conditions the phase signal of interest $\varphi$ has been effectively amplified. Such a signal amplification through time-reversed interaction (SATIN) protocol can make use of complex states with large quantum Fisher information, that are not necessarily simple squeezed states with a Gaussian envelope, and can potentially provide HS and sensitivity quite close to the HL even at limited resolution of the final measurement \cite{Davis2016,Frowis2016,Nolan2017}.

Previously, non-Gaussian many-body entangled states have been experimentally generated in Bose-Einstein condensates \cite{Strobel2014,Lucke2016}, neutral cold atoms~\cite{McConnell-Vuletic2015,barontini2015Zeno}, and cold trapped ions \cite{Bohnet2016}, while time-reversal-type protocols have been implemented using phase shifts in a three-level system for neutral atoms \cite{Linnemann2016}, and using the coupling to a motional mode in combination with spin rotations for trapped ions \cite{gilmore2021quantum}. Experiments demonstrating HS have also been performed, that have, however, either used squeezed spin states (SSSs) and therefore been limited far ($>46$~dB) from the HL \cite{Bohnet2014a}, or have involved a relatively small number of atoms $N \leq 20$ \cite{Monz-Blatt2011,omran2019CatState}.  Furthermore, amplification of the quantum phase in a neutral atom system coupled to an optical resonator has been demonstrated through a protocol interspersing spin squeezing with a state rotation. Remarkably, this has enabled the detection of -8 dB noise reduction without the need of detection resolution below the SQL \cite{Hosten2016a}. This protocol requires, and has been performed with, Gaussian states.

Here, following the SATIN protocol proposed in Ref. \cite{Davis2016}, we create a highly non-Gaussian entangled state in a system of $^{171}$Yb atoms and demonstrate phase sensitivity with HS ($b/N$) at fixed distance $b=12.6$~dB from the HL. When used in a Ramsey sequence in an atomic interferometer, we achieve the highest metrological gain over the SQL, $\mathcal{G} = 11.8 \pm 0.5$~dB, that has been achieved in any (full Ramsey) interferometer to date, and comparable to the gain $\mathcal{G} =10.5\pm 0.3 $~dB achieved with much larger atom number $N=1\times 10^5$~\cite{Hosten2016}.

\begin{figure}[hbtp]
\setlength{\unitlength}{1\textwidth}
\includegraphics[width=163mm,scale=1]{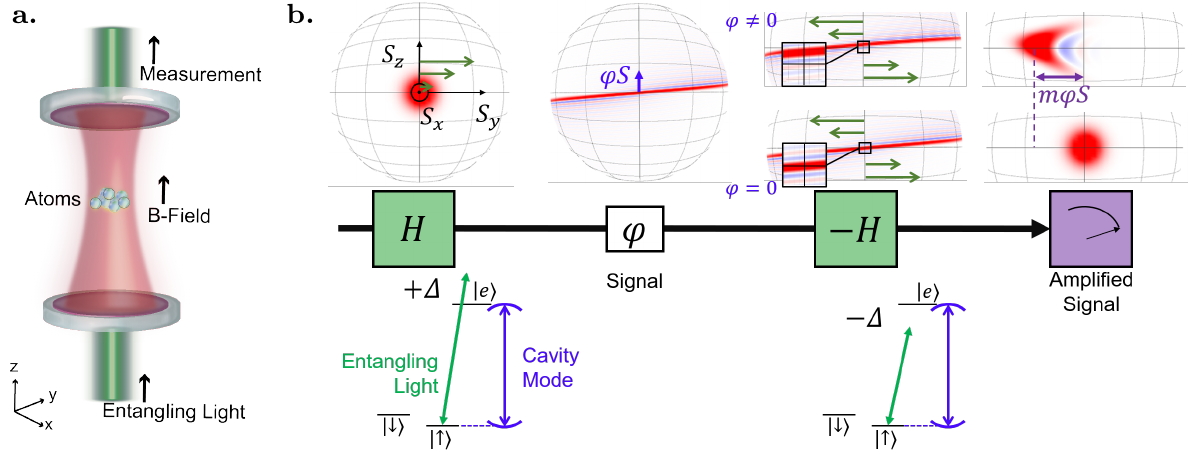}
\caption{
\textbf{Setup and sequence.} \textbf{a.} $^{171}$Yb atoms are trapped inside an optical cavity in an optical lattice. Light for state preparation, entanglement generation, and state measurement (green) is sent through the cavity along the $z$-axis. The static magnetic field defines the quantization axis parallel to $z$. \textbf{b.} SATIN protocol sequence with the quantum-state evolution (top) and relevant energy-levels and cavity mode (bottom). Wigner quasiprobability distribution functions describing the collective quantum state are calculated for an ensemble of 220 atoms, and are represented on the generalized Bloch sphere for the ground-state manifold $\{\ket{\downarrow},\ket{\uparrow}\}$. An entangling light pulse is passed through the cavity detuned by $+\Delta$ from the $\ket{\uparrow}\rightarrow\ket{e}$ transition and the cavity mode. This light generates the nonlinear OAT Hamiltonian $H\propto S_z^2$ which shears the initial CSS state (green arrows on the generalized Bloch sphere). A rotation $\varphi$ about $ S_y$ displaces the state by $\Delta S_z= \varphi S$. Subsequently, a (dis)entangling light pulse is sent through the cavity detuned by $-\Delta$ from the $\ket{\uparrow}\rightarrow\ket{e}$ transition and the cavity mode. This pulse generate the negative OAT Hamiltonian $-H$ which causes the quantum state to evolve effectively "backwards in time". With $\varphi=0$ the quantum state evolves back to the original CSS, while for a small angle ($\varphi\neq0$) the final state is displaced by an angle $m\varphi$ from the original CSS, where $m$ is the SATIN signal amplification.
}
\label{fig:Scheme}
\end{figure}

Our system consists of up to $N=400$ laser cooled $^{171}$Yb atoms that are trapped in an optical lattice inside a high-finesse optical resonator (Fig. 1a) \cite{braverman2019near}. We work in the nuclear-spin manifold $s=\frac{1}{2}$ of the electronic ground state $^1S_0$, and first create a state of the collective atomic spin ${\bf S}=\sum {\bf s}_i$ pointing along the $x$-axis (coherent spin state, CSS). In this product state of the individual spins ${\bf s}_i$, each atom is in a superposition of the  states $\ket{\uparrow} \equiv \ket{m_s=+\frac{1}{2}}$ and $\ket{\downarrow} \equiv \ket{m_s=-\frac{1}{2}}$. We then apply the one-axis twisting (OAT) Hamiltonian~\cite{Kitagawa1993}
\begin{equation}
H=\chi S_z^2 
\label{eq:Hamiltonian}
\end{equation}
to the CSS to create an entangled state (see Fig. 1). The OAT Hamiltonian is generated by the nonlinear interaction between the atoms and light that is being applied to the cavity \cite{braverman2019near,Pedrozo2020entanglement} (see Supplementary Materials). The light-atom interaction as a generator of entanglement offers the advantage that it can not only be turned on or off arbitrarily, but also that the sign of $H$ can be changed by adjusting the frequency of the incident light (see Supplementary Materials). 
When we apply $H$ for a time $t$, the state evolves under the OAT operator 
$
    \hat{U} = \exp \left( -i\frac{\tilde{Q}}{\sqrt{N}}S_z^2 \right),
$
where we have introduced the normalized twisting parameter $\tilde{Q} \equiv \sqrt{N}\chi t$. (Here $\tilde{Q}=2\pi$ would correspond to a state wrapped all around the Bloch sphere.

\begin{figure}[hbtp]
\setlength{\unitlength}{1\textwidth}
\includegraphics[width=100mm,scale=1]{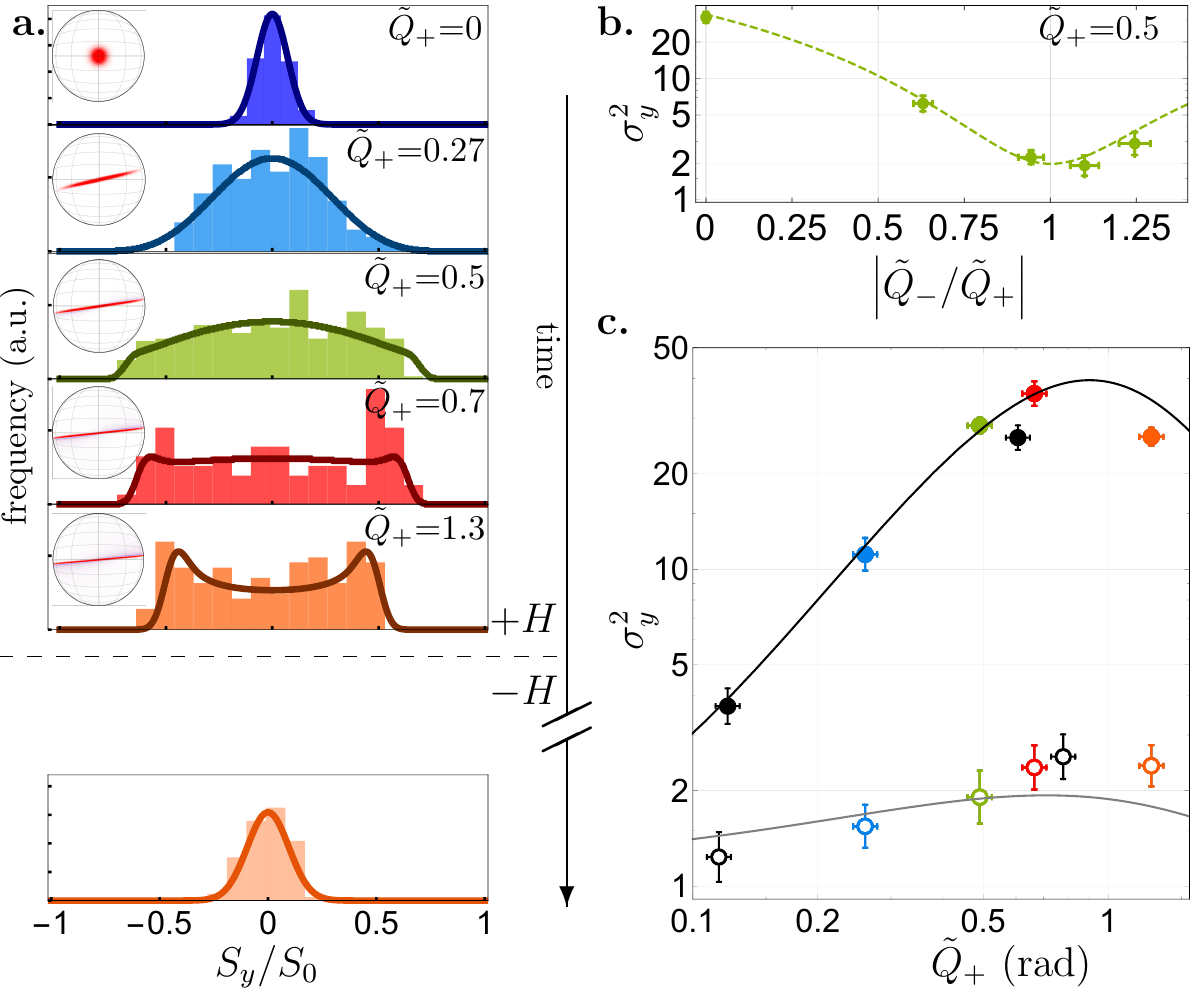}
\caption{
\textbf{State evolution during SATIN sequence.} \textbf{a.} Histograms showing the distributions of $S_y$ measured at different times during the SATIN sequence. The system transitions from Gaussian states (blue histograms) to strongly non-Gaussian states (red, orange). After reversing the sign of the Hamiltonian, the original CSS is almost recovered (orange, lower plot). The solid lines are the predicted distributions from a theoretical model including finite contrast $\mathcal{C}$, measurement resolution $\sigma^2_{meas}=0.15$, and excess broadening from light-atom entanglement $\mathcal{I}$. The Bloch spheres show the expected Wigner quasi-probability distributions at the corresponding times. 
\textbf{b.} The normalized $S_y$ variance as a function of the unshearing strength $\tilde{Q}_{-}$ for $\tilde{Q}_{+}=0.5$. The dashed line shows the model prediction. 
\textbf{c.} Measured variances of $S_y$ resulting from shearing with $\tilde{Q}_{+}$  (filled circles), and after the corresponding unshearing $\tilde{Q}_{-}$ (filled diamonds). Solid lines are theoretical predictions, including $\mathcal{C}$, $\sigma^2_{meas}$ and $\mathcal{I}$. All error bars represent 1$\sigma$ statistical uncertainty resulting from 100 to 150 experimental realizations. Error bars of $\sigma_y^2$ are inferred by bootstrapping the data.}
\label{fig:2}
\end{figure}

We first characterize the action of the OAT Hamiltonian $H$ and the effective evolution "backwards in time" that can be obtained by applying $-H$ subsequently to $H$. To this end we measure for various twisting strengths $\tilde{Q}_{+}$ the $S_y$ spin distribution and its normalized variance $\sigma_y^2 \equiv 2(\Delta S_y)^2/S_0$. (Here $S_0=N/2$ and the SQL, obtained for the CSS, corresponds to $\sigma_y^2=1$).
For small $\tilde{Q}_{+} \ll 1$, the OAT operator $\hat{U}(\tilde{Q}_{+})$  creates an SSS with a Gaussian envelope, while for $\tilde{Q}_{+}\geq0.5$ the state stretches around a substantial portion of the Bloch sphere. As Fig. 2a shows, we observe strongly non-Gaussian probability distributions for $S_y$ that agree well with the expected evolution from the OAT Hamiltonian calculated without any free parameters (see Supplementary Materials). We also verify that the $S_z$ distribution remains unaffected by the OAT.

If we subsequently apply the negative Hamiltonian and the corresponding untwisting operator $\hat{U}(\tilde{Q}_{-})$, then for a matched untwisting magnitude, $\tilde{Q}_{-}=-\tilde{Q}_{+}$, the state distribution along $S_y$ reverts back to a Gaussian distribution with a variance that is only slightly increased compared to the original CSS (Fig. 2a bottom and 2b). The residual broadening can be explained by the fact that the OAT Hamiltonian $H$ of Eq. \ref{eq:Hamiltonian}, is only an approximation to the actual physical process, where the transmitted and scattered light carries some residual information about the atomic spin $S_z$. Then tracing  over the unobserved light degrees of freedom causes an excess broadening of $\sigma_y^2$ by a factor $1+\mathcal{I}$ \cite{braverman2019near}. To quantify the excess broadening $\mathcal{I}$, we fix $\tilde{Q}_{+}$ and measure $\sigma_y^2$ vs. $\tilde{Q}_{-}$. 
Data for $\tilde{Q}_{+}=0.5$ and $N {=} 220\pm{20}$ atoms are shown in Fig. 2b. It is clear that $\sigma_y^2$ is indeed minimized near $\tilde{Q}_{-} = -\tilde{Q}_{+}$, with a small excess broadening of $\mathcal{I}= 0.9\pm 0.4$ after accounting for measurement resolution (increasing $\sigma_y^2$ by $0.15\pm0.02$) and contrast loss (decreasing $\sigma_y^2$ by $0.7 \pm 0.1$). Our algebraic model, without any fitting parameters, agrees remarkably well with the data. Fig. 2c shows the variance of $S_y$ resulting from the shearing $\tilde{Q}_{+}$ (solid circles), and after the unshearing $\tilde{Q}_{-}$ (open circles). The data points are fitted to the theoretical curve taking into account the different sources of decoherence: excess broadening $\mathcal{I}$, finite contrast $\mathcal{C}$, and measurement resolution $\sigma^2_{meas}$. 

\begin{figure}[hbtp]
\setlength{\unitlength}{1\textwidth}
\includegraphics[width=100mm,scale=.95]{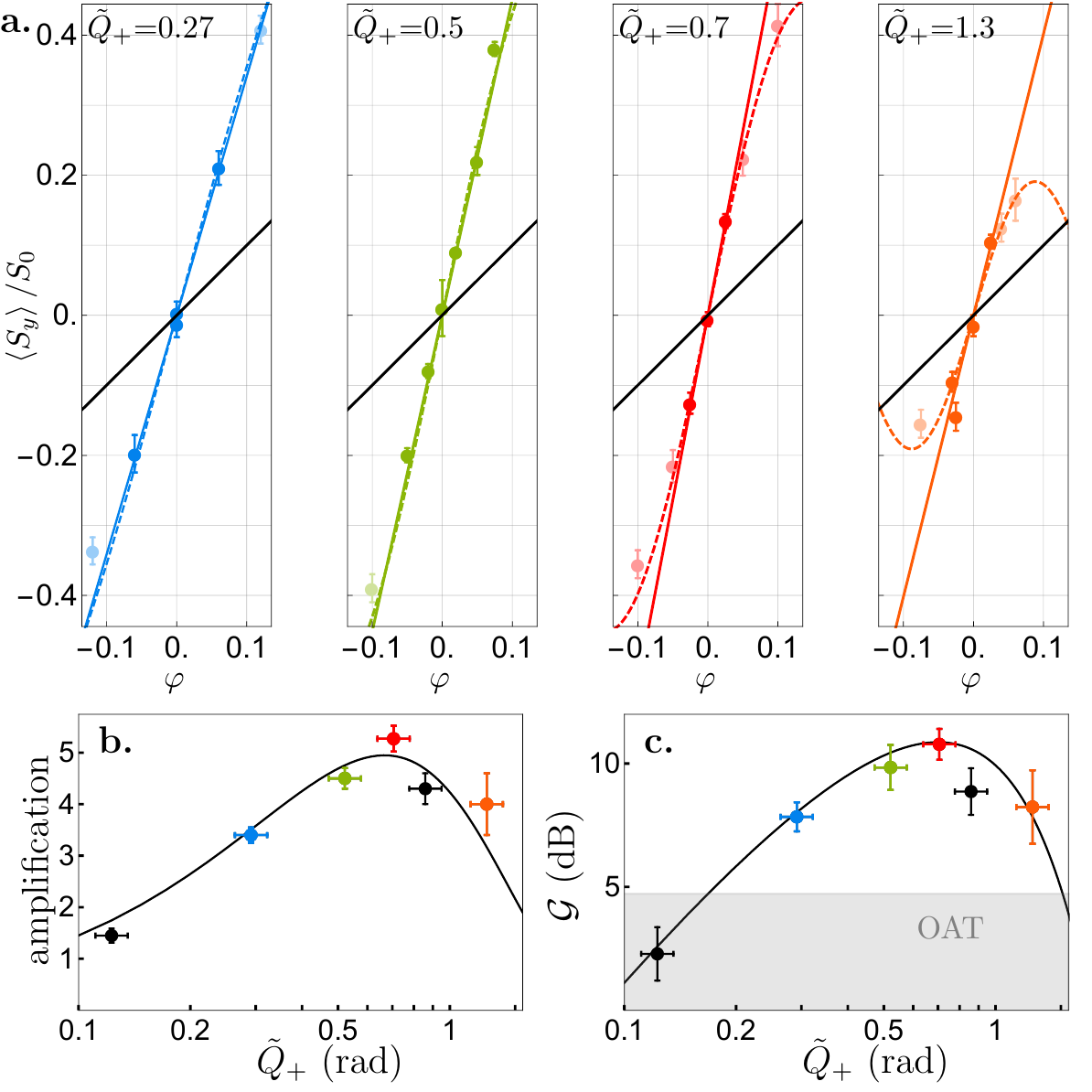}
\caption{
\textbf{Signal amplification and metrological gain.} All data are taken with $N=220\pm12$ atoms. a) The mean value  $\langle S_y\rangle/S_0$ as a function of the angle $\varphi$. The solid black line is the maximal signal that can be reached with a CSS. Dashed lines represent calculations, solid lines are the linear fit to the data for small $\varphi$. \textbf{b.} The signal amplification as a function of the shearing parameter $\tilde{Q}_{+}$. Solid line is the model prediction. \textbf{c.} The resulting metrological gain $\mathcal{G}$ as a function of the squeezing strength $\tilde{Q}_{+}$, solid lines show the model prediction. The gray area represents the metrological gain accessible with a simple OAT squeezing in the same system with a measurement resolution $\sigma^2_{meas}=0.15$ in variance ($-8.2~$dB). Error bars represent 1$\sigma$ confidence intervals. }
\label{fig:fig3}
\end{figure}

We next measure the small-signal amplification $m$ provided by the SATIN protocol. To this end, we prepare a strongly entangled state by evolving a CSS under the OAT-operator $\hat{U}(\tilde{Q}_{+})$, rotate this state by a small angle $\varphi$ around the $y$-axis such that $\left< S_z\right> = \varphi S_0$, and apply the untwisting operator $\hat{U}(\tilde{Q}_{-}=-\tilde{Q}_{+})$ which amplifies $\varphi$ by a factor $m$ and maps it onto the $y$-axis, resulting in $\left<S_y\right>=m\varphi S_0$.
We measure how $\left<S_y\right>/S_0$ scales with $\varphi$ for different $\tilde{Q}_{+}$ and present the results in Fig. 3a, which compares the signal amplification of the SATIN scheme to a measurement with a CSS (for which $m_{\mathrm{SQL}} = 1$).
Note that for large displacements $\varphi$ the finite size of the Bloch sphere makes the mapping of the rotation angle $\varphi$ onto $\aver{S_y}$ nonlinear. The measured data are well described by the model (see Supplementary Materials).

Fig. 3b shows the amplification $m$ vs. $\tilde{Q}_{+}$ together with the theoretical model.
The amplification scales linearly with $\tilde{Q}_{+}$ for $\tilde{Q}_{+} \ll 1$ and reaches its maximum at $\tilde{Q}_{+} \approx 0.7$, larger than the optimum value $\tilde{Q}_{+}=0.4$ for minimizing the variance of the spin squeezed state for the same atom number \cite{braverman2019near}. Even the state with $\tilde{Q}_{+}=1.3$ outperforms the squeezed state by several dB. This non-Gaussian state has as root-mean-square (rms) phase spread of $1.3$~rad, where a state with a uniform distribution between 0 and $2\pi$ would have an rms phase spread of $\pi/\sqrt{3}\approx1.8$~rad.
This demonstrates the usefulness of non-Gaussian states for quantum metrology.

The sensitivity $\delta \varphi$ of the SATIN protocol, i.e., the minimal resolvable displacement of a state, can be estimated as the displacement $m S_0 \delta \varphi$ at the end of the sequence that equals the measured uncertainty $\Delta S_y= \sigma_y \sqrt{S_0/2}$ after the twisting-untwisting sequence for $\varphi=0$.
Thus, the gain of the SATIN protocol over the SQL with sensitivity $\left( \Delta\varphi \right)_\mathrm{SQL}= 1/\sqrt{2S_0}$ is given by 
\begin{equation}
    \mathcal{G} =  \frac{\left( \Delta\varphi \right)^2_\mathrm{SQL}}{\left( \delta\varphi\right)^2} = \frac{m^2}{\sigma_y^2} = \frac{S_0}{2}\frac{m^2}{(\Delta S_y)^2}.
\end{equation}
In Fig.~3c we present the resulting metrological gain $\mathcal{G}(\tilde{Q}_{+})$ for an ensemble of $N=220$ atoms. 
Since $\sigma_y^2$ does not change significantly when increasing  $\tilde{Q}_{+}$, the metrological gain approximately follows the behavior of $m$ and peaks around $\tilde{Q}_{+} \approx 0.7$.
At larger values of $\tilde{Q}_{+}$ the signal amplitude $\langle S_y\rangle$ is reduced due to contrast loss from photon scattering into free space. 
For $N=220$ atoms the metrological gain peaks at $\mathcal{G}=10.8\pm0.6$~dB. 
This is more than $6$~dB larger than the maximal gain achievable with spin squeezing in the same system, which is limited by measurement resolution and decoherence to $4.7$~dB.

Furthermore, we investigate the scaling of the sensitivity with atom number $N$. Unlike spin squeezing \cite{Kitagawa1993} and the quantum magnification protocol of Ref. \cite{Hosten2016a}, the SATIN protocol is not limited by the curvature of the Bloch sphere, and therefore we expect that under optimum conditions the precision improves in proportion to the atom number, corresponding to HS.
Fig. 4 shows that when we vary $N$ between 50 and 370, we indeed measure a gain over the SQL that varies as $\mathcal{G} \propto N$, achieving HS. This implies that the averaging time necessary to achieve a certain resolution improves as $N^2$ for the SATIN protocol, as predicted in Ref. \cite{Davis2016}. In particular, for $N = 370 \pm 20$ atoms we reach $\mathcal{G} = 12.8 \pm 0.9$~dB.

\begin{figure}[hbtp]
\setlength{\unitlength}{1\textwidth}
\includegraphics[width=100mm,scale=.9]{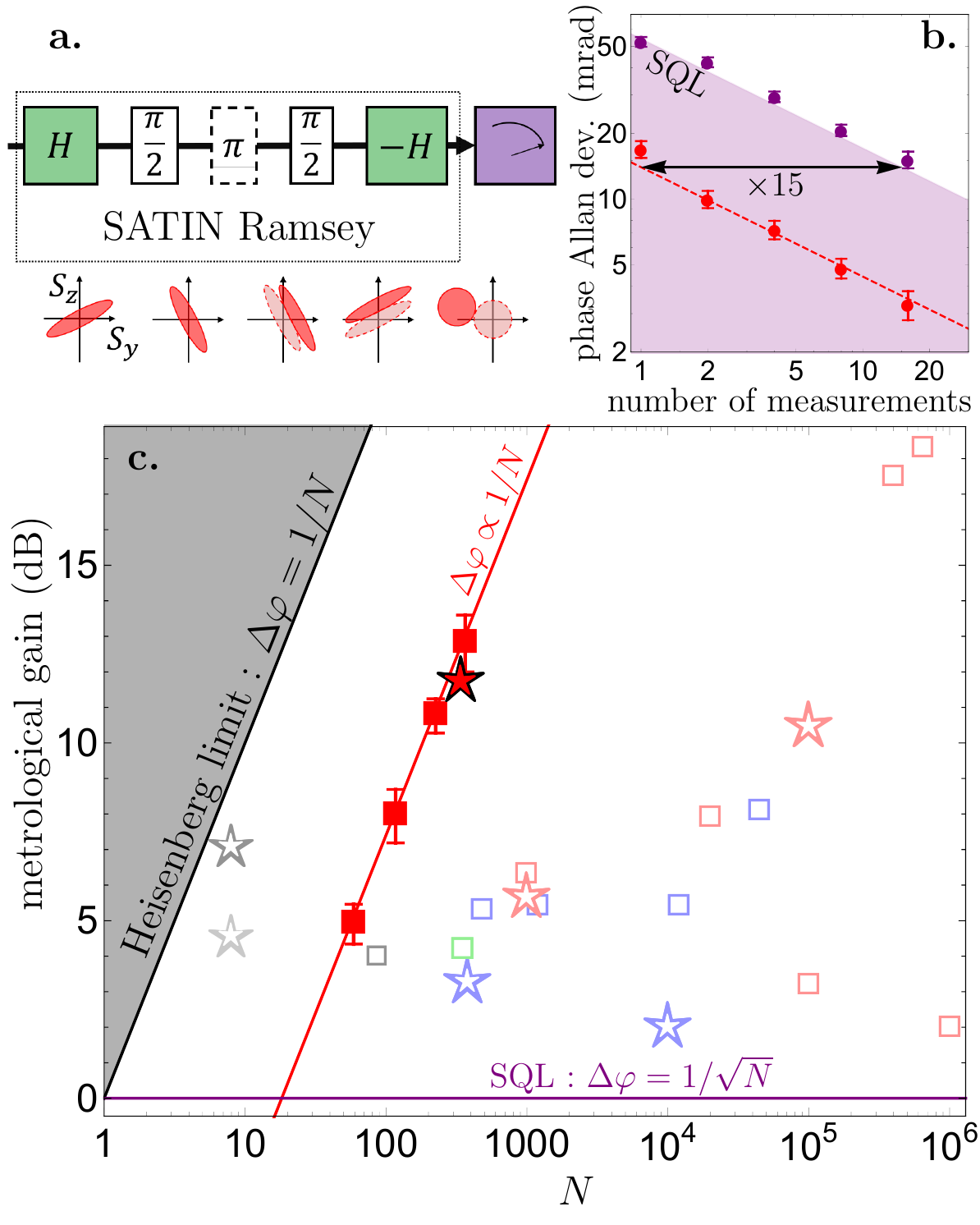}
\caption{
\textbf{Scaling of sensitivity with atom number and comparison with previous results.}  \textbf{a.} Block diagram showing a SATIN scheme applied to a Ramsey interferometry experiment. \textbf{b.} Phase Allan deviation plot of Ramsey spin-echo interferometry performed with a CSS (purple circles) and an optimally over-squeezed state in a SATIN protocol (red circles). In both cases we used $340{\pm}20$ atoms in the interferometer. The shaded purple area indicates the region below the SQL. The data for the SATIN sequence is fitted to a white noise model (red dashed line) showing $11.8{\pm}0.5$~dB of metrological gain over the SQL. \textbf{c.} Comparison with previews results. Blue data: BEC experiments~\cite{Hamley2012,berrada2013integrated,muessel2014Scalable, kruse2016improvement, schmied2016bell}. Red data: Thermal atoms experiments~\cite{Appel2009,Sewell2012,Cox2016a,Hosten2016,braverman2019near, huang2020self}. Black data: Ions~\cite{Monz-Blatt2011, Bohnet2016}. Gray data: Rydberg atoms in a tweezer array~\cite{omran2019CatState}.
Green data: squeezing generated in an optical-lattice clock~\cite{Pedrozo2020entanglement}. Squares are expected metrological gains, obtained by quantum state characterization. Stars refer to directly measured phase-sensitivity gain. Filled symbols are obtained in this work.
Errorbars represent 1$\sigma$ confidence interval.}
\label{fig:fig4}
\end{figure}

Finally, we implement a full (phase) interferometer for ac magnetic fields in the form of a spin echo SATIN Ramsey sequence. We first apply the OAT operator, then rotate the oversqueezed state about $S_x$ with $\pi/2$-pulse, thus making it sensitive to phase perturbations (i.e. rotations around $S_z$). Subsequently, we rotate the state back by another $\pi/2$ pulse about $S_x$ before subjecting it to the negative Hamiltonian. To cancel fluctuations of the static magnetic field, we apply a spin echo $\pi$-pulse, separated from the $\pi/2$ pulses by $1.73$~ms, thus realizing an interferometer sensitive to ac magnetic fields with peak sensitivity at $290$~Hz.    
We observe a metrological gain of  $\mathcal{G} = 11.8 \pm 0.5$~dB with $N=340{\pm}20$ atoms in the interferometer (solid red star in Fig.~4c.), slightly exceeding the previous record of $10.5\pm0.3$~dB \cite{Hosten2016} in a Ramsey interferometer (atomic clock) with large atom number. The gain $\mathcal{G}$ achieved with the SATIN Ramsey sequence represents a factor of 15 of reduction in averaging time for a  desired precision (see Fig. 4b). Our Ramsey interferometer also performs near the HS limit for the SATIN protocol, as shown by the data point (filled red star) in Fig. 4c.

Under ideal conditions, the SATIN protocol provides a metrological gain only 4.3~dB away from the HL~\cite{Davis2016} for an optimized shearing strength $\tilde{Q}^{opt}_{+}=1$. 
Dissipation in the atomic system, in our system due to photon scattering and light-atom entanglement, reduces the maximum available gain and the optimum shearing parameter $\tilde{Q}^{opt}_+$. For our parameters, $\tilde{Q}^{opt}_{+}=0.7$, which reduces the metrological gain by $0.9$~dB, while the interferometer contrast loss and non-unitary state evolution under the full Hamiltonian contribute $4.4$~dB and $3.2$~dB, respectively.
The joint effect of these imperfections imposes a distance $12.6~$dB from the HL. To move closer to the HL, the dissipation in the system must be reduced by increasing the atom-cavity coupling, as characterized by the single-atom cooperativity $\eta$ (see Supplementary Materials).
For example, by increasing $\eta$ by a factor 10 to $\eta=80$, we expect performance only $8$~dB away from the HL.
At present, we have seen no deviation from HS, i.e. the measurement precision improves linearly with atom number $N$. The latter can likely be increased in the future by means of optimized loading protocols, like a recooling and retrapping sequence~\cite{hu2017creation}, in the two-dimensional optical lattice. 

There is no fundamental limitation to achieve HS with larger atom numbers $N$ in our system, but technical parameters, such as the squeezing light and of the RF rotation pulses, need to be stringently controlled. The latter control has already been successfully demonstrated in systems of up to $\sim 10^6$ atoms \cite{Cox2016a, Hosten2016}. The frequency of the squeezing light is stabilized to an ultralow-thermal-expansion cavity so that frequency noise will not affect the performance even for much larger $N$. The relative energy of the $Q_+$/$Q_-$ shearing/unshearing pulses needs to be controlled near the shot noise level, which corresponds to a moderate 5\% for our ensemble size, and thus light intensity noise can also be controlled sufficiently well to maintain the HS for larger systems.

Our protocol can be used for a variety of fundamental and applied purposes, such as tests of fundamental laws of physics \cite{Safronova2018RevModPhys,Safronova2019}, the search for physics beyond the Standard Model \cite{Pospelov2013GNOME,Derevianko2014DarkMatter,Arvanitaki2015DarkMatter,Wcislo2018}, the detection of gravitational waves \cite{Kolkowitz2016}, or geodesy \cite{lisdat2016clock,grotti2018geodesy,Katori2020}. As with all entanglement-based protocols beyond the SQL, they are useful to boost the sensitivity in applications that require performance in a given bandwidth or limited time. For instance, by coherently transferring the entanglement onto the ultra-narrow optical clock transition by means of a high-fidelity optical $\pi$-pulse \cite{Pedrozo2020entanglement,young2020half}, the protocol can be directly used to search for transient changes in the fundamental constants induced by dark matter \cite{Derevianko2014DarkMatter}.

\section*{Acknowledgments}
We thank Boris Braverman, Akio Kawasaki, Mikhail Lukin, and Jun Ye for discussions. 
\paragraph*{Fundings:}This work was supported by NSF (grant no. PHY-1806765), DARPA (grant no. D18AC00037), ONR (grant no. N00014-20-1-2428), the NSF Center for Ultracold Atoms (CUA) (grant no. PHY-1734011), and NSF QLCI-CI QSEnSE (grant no. 2016244). S.C. and A.A. acknowledge support from the Swiss National Science Foundation (SNSF).
\paragraph*{Author contributions:}
S.C., E.P.-P., A.A., and Z.L. led the experimental efforts and simulations. S.C., E.P.-P., A.A., and Z.L. contributed to the data analysis. V.V. conceived and supervised the experiment. S.C., E.P.-P., A.A., and V.V. wrote the manuscript. All authors discussed the experiment implementation, the results, and contributed to the manuscript.

\bibliography{SqUnsq.bib}

\begin{thebibliography}{10}

\bibitem{Takano2009}
T.~Takano, M.~Fuyama, R.~Namiki, Y.~Takahashi, {\it Phys. Rev. Lett.\/} {\bf
  102}, 033601 (2009).

\bibitem{Appel2009}
J.~Appel, {\it et~al.\/}, {\it Proc. Natl. Acad. Sci. U.S.A.\/} {\bf 106},
  10960 (2009).

\bibitem{Sewell2012}
R.~J. Sewell, {\it et~al.\/}, {\it Phys. Rev. Lett.\/} {\bf 109}, 253605
  (2012).

\bibitem{Hamley2012}
C.~D. Hamley, C.~Gerving, T.~Hoang, E.~Bookjans, M.~S. Chapman, {\it Nat.
  Phys.\/} {\bf 8}, 305 (2012).

\bibitem{berrada2013integrated}
T.~Berrada, {\it et~al.\/}, {\it Nature communications\/} {\bf 4}, 1 (2013).

\bibitem{muessel2014Scalable}
W.~Muessel, H.~Strobel, D.~Linnemann, D.~B. Hume, M.~K. Oberthaler, {\it Phys.
  Rev. Lett.\/} {\bf 113}, 103004 (2014).

\bibitem{schmied2016bell}
R.~Schmied, {\it et~al.\/}, {\it Science\/} {\bf 352}, 441 (2016).

\bibitem{Cox2016a}
K.~C. Cox, G.~P. Greve, J.~M. Weiner, J.~K. Thompson, {\it Phys. Rev. Lett.\/}
  {\bf 116}, 093602 (2016).

\bibitem{Hosten2016}
O.~Hosten, N.~J. Engelsen, R.~Krishnakumar, M.~A. Kasevich, {\it Nature
  (London)\/} {\bf 529}, 505 (2016).

\bibitem{Hosten2016a}
O.~Hosten, R.~Krishnakumar, N.~J. Engelsen, M.~A. Kasevich, {\it Science\/}
  {\bf 352}, 1552 (2016).

\bibitem{Bohnet2016}
J.~G. Bohnet, {\it et~al.\/}, {\it Science\/} {\bf 352}, 1297 (2016).

\bibitem{braverman2019near}
B.~Braverman, {\it et~al.\/}, {\it Phys. Rev. Lett.\/} {\bf 122}, 223203
  (2019).

\bibitem{Bao2020}
H.~Bao, {\it et~al.\/}, {\it Nature (London)\/} {\bf 581}, 159 (2020).

\bibitem{Pedrozo2020entanglement}
E.~Pedrozo-Pe\~{n}afiel, {\it et~al.\/}, {\it Nature (London)\/} {\bf 588}, 414
  (2020).

\bibitem{huang2020self}
M.-Z. Huang, {\it et~al.\/}, {\it arXiv preprint arXiv:2007.01964\/}  (2020).

\bibitem{kruse2016improvement}
I.~Kruse, {\it et~al.\/}, {\it Physical review letters\/} {\bf 117}, 143004
  (2016).

\bibitem{Kitagawa1993}
M.~Kitagawa, M.~Ueda, {\it Phys. Rev. A\/} {\bf 47}, 5138 (1993).

\bibitem{Wineland1994}
D.~J. Wineland, J.~J. Bollinger, W.~M. Itano, D.~J. Heinzen, {\it Phys. Rev.
  A\/} {\bf 50}, 67 (1994).

\bibitem{Pezze2018}
L.~Pezz{\`e}, A.~Smerzi, M.~K. Oberthaler, R.~Schmied, P.~Treutlein, {\it Rev.
  Mod. Phys.\/} {\bf 90}, 035005 (2018).

\bibitem{Leibfried2005}
D.~Leibfried, {\it et~al.\/}, {\it Nature (London)\/} {\bf 438}, 639 (2005).

\bibitem{Monz-Blatt2011}
T.~Monz, {\it et~al.\/}, {\it Phys. Rev. Lett.\/} {\bf 106}, 130506 (2011).

\bibitem{omran2019CatState}
A.~Omran, {\it et~al.\/}, {\it Science\/} {\bf 365}, 570 (2019).

\bibitem{DiCarlo-CatSupercond}
L.~DiCarlo, {\it et~al.\/}, {\it Nature (London)\/} {\bf 467}, 574 (2010).

\bibitem{Song-10CatSuperCond}
C.~Song, {\it et~al.\/}, {\it Phys. Rev. Lett.\/} {\bf 119}, 180511 (2017).

\bibitem{Saffman2009}
M.~Saffman, D.~Oblak, J.~Appel, E.~S. Polzik, {\it Phys. Rev. A\/} {\bf 79},
  023831 (2009).

\bibitem{Davis2016}
E.~Davis, G.~Bentsen, M.~Schleier-Smith, {\it Phys. Rev. Lett.\/} {\bf 116},
  053601 (2016).

\bibitem{Frowis2016}
F.~Fr\"{o}wis, P.~Sekatski, W.~D\"{u}r, {\it Phys. Rev. Lett.\/} {\bf 116},
  090801 (2016).

\bibitem{Toscano2006}
F.~Toscano, D.~A.~R. Dalvit, L.~Davidovich, W.~H. Zurek, {\it Phys. Rev. A\/}
  {\bf 73}, 023803 (2006).

\bibitem{Nolan2017}
S.~P. Nolan, S.~S. Szigeti, S.~A. Haine, {\it Phys. Rev. Lett.\/} {\bf 119},
  193601 (2017).

\bibitem{Macri2016}
T.~Macr\`{i}, A.~Smerzi, L.~Pezz\`{e}, {\it Phys. Rev. A\/} {\bf 94}, 010102
  (2016).

\bibitem{Strobel2014}
H.~Strobel, {\it et~al.\/}, {\it Science\/} {\bf 345}, 424 (2014).

\bibitem{Lucke2016}
B.~L\"{u}cke, {\it et~al.\/}, {\it Science\/} {\bf 334}, 773 (2016).

\bibitem{McConnell-Vuletic2015}
R.~McConnell, H.~Zhang, J.~Hu, S.~Ćuk, V.~Vuleti\'{c}, {\it Nature (London)\/}
  {\bf 519}, 439 (2015).

\bibitem{barontini2015Zeno}
G.~Barontini, L.~Hohmann, F.~Haas, J.~Est{\`e}ve, J.~Reichel, {\it Science\/}
  {\bf 349}, 1317 (2015).

\bibitem{Linnemann2016}
D.~Linnemann, {\it et~al.\/}, {\it Phys. Rev. Lett.\/} {\bf 117}, 013001
  (2016).

\bibitem{gilmore2021quantum}
K.~A. Gilmore, {\it et~al.\/}, {\it Science\/} {\bf 373}, 673 (2021).

\bibitem{Bohnet2014a}
J.~G. Bohnet, {\it et~al.\/}, {\it Nat. Photonics\/} {\bf 8}, 731 (2014).

\bibitem{hu2017creation}
J.~Hu, {\it et~al.\/}, {\it Science\/} {\bf 358}, 1078 (2017).

\bibitem{Safronova2018RevModPhys}
M.~S. Safronova, {\it et~al.\/}, {\it Rev. Mod. Phys.\/} {\bf 90}, 025008
  (2018).

\bibitem{Safronova2019}
M.~S. Safronova, {\it Annalen der Physik\/} {\bf 531}, 1800364 (2019).

\bibitem{Pospelov2013GNOME}
M.~Pospelov, {\it et~al.\/}, {\it Phys. Rev. Lett.\/} {\bf 110}, 021803 (2013).

\bibitem{Derevianko2014DarkMatter}
A.~Derevianko, M.~Pospelov, {\it Nature Physics\/} {\bf 10}, 933 (2014).

\bibitem{Arvanitaki2015DarkMatter}
A.~Arvanitaki, J.~Huang, K.~Van~Tilburg, {\it Phys. Rev. D\/} {\bf 91}, 015015
  (2015).

\bibitem{Wcislo2018}
P.~Wcis{\l}o, {\it et~al.\/}, {\it Science Advances\/} {\bf 4} (2018).

\bibitem{Kolkowitz2016}
S.~Kolkowitz, {\it et~al.\/}, {\it Phys. Rev. D\/} {\bf 94}, 124043 (2016).

\bibitem{lisdat2016clock}
C.~Lisdat, {\it et~al.\/}, {\it Nature communications\/} {\bf 7}, 1 (2016).

\bibitem{grotti2018geodesy}
J.~Grotti, {\it et~al.\/}, {\it Nature Physics\/} {\bf 14}, 437 (2018).

\bibitem{Katori2020}
M.~Takamoto, {\it et~al.\/}, {\it Nature Photonics\/} {\bf 14}, 411 (2020).

\bibitem{young2020half}
A.~W. Young, {\it et~al.\/}, {\it Nature\/} {\bf 588}, 408 (2020).

\bibitem{Lee2014a}
J.~Lee, G.~Vrijsen, I.~Teper, O.~Hosten, M.~A. Kasevich, {\it Opt. Lett.\/}
  {\bf 39}, 4005 (2014).

\bibitem{Kawasaki2019}
A.~Kawasaki, {\it et~al.\/}, {\it Phys. Rev. A\/} {\bf 99}, 013437 (2019).

\bibitem{li2021collective}
Z.~Li, {\it et~al.\/}, {\it arXiv preprint arXiv:2106.13234\/}  (2021).

\bibitem{schulte2020ramsey}
M.~Schulte, V.~J. Mart{\'\i}nez-Lahuerta, M.~S. Scharnagl, K.~Hammerer, {\it
  Quantum\/} {\bf 4}, 268 (2020).

\bibitem{koczor2020fast}
B.~Koczor, R.~Zeier, S.~J. Glaser, {\it Physical Review A\/} {\bf 102}, 062421
  (2020).

\end{thebibliography}

\bibliographystyle{Science}

\section*{Supplementary Materials}

\paragraph*{Atom loading and cooling}

We load $^{171}$Yb atoms into a two-color mirror magneto-optical trap (MOT) on the singlet ${^1S_0}{\rightarrow} {^1P_1}$ and triplet ${^1S_0}{\rightarrow} {^3P_1}$ transitions, followed by a second-stage green MOT on the triplet transition. 
By changing the magnetic field, the atomic cloud is then transported into the intersection region of the cavity TEM$_{00}$ mode and a one-dimensional optical lattice along the $x$-direction. 
The trap is formed by `magic-wavelength' light with $\lambda_t\approx759$ nm, and the trap depth is $U_x{=}k_B{\times}10$~$\mu$K. 
The green MOT light is then turned off, and the magic-wavelength trap inside the cavity, detuned from the $x$ lattice by $160$~MHz to avoid interference, is ramped up in 40 ms to a trap depth $U_{c}{=}k_B{\times}120$~$\mu$K. 
At the end of loading process, the transverse lattice power is ramped down to zero and back to full power in 50 ms to remove all the atoms that are outside the overlap region of the two lattices. 
In this way, an ensemble of atoms is prepared at a distance of $180$~$\mu$m from the end mirror of the cavity, where the single-atom peak cooperativity is $\eta=7.7\pm0.3$.

After loading, Raman sideband cooling is performed on the transition ${^1S_0}{\rightarrow} {^3P_1}$ in an applied magnetic field $B_z=13.6$~G along the $z$-direction. 
In $100$~ms, the atomic temperature is lowered to $\approx2$~$\mu$K, corresponding to an average motional occupation number $\aver{n_x}{=}0.2$ at a trap vibration frequency of $\omega_x/(2\pi)=67$~kHz along the $x$-direction. 
The cavity trap is then adiabatically ramped down to $U_c{=}k_B{\times}40$~$\mu$K to further reduce temperature.
We observe that during the Raman sideband cooling, where the optical pumping is provided by intracavity light, the atoms reorganize along the lattice such that all atoms have nearly the maximum coupling $\eta$ to the cavity mode and the squeezing light. Previously, this has been achieved with a wavelength of the trapping light that was twice the probing wavelength \cite{Lee2014a}. Here, contemplating applications on the optical-clock transition \cite{Pedrozo2020entanglement}, our trap is at a magic wavelength for the clock transition.

\paragraph*{Initialization of the experimental sequence}

\begin{figure}[hbtp]
\setlength{\unitlength}{1\textwidth}
\includegraphics[width=160mm,scale=.9]{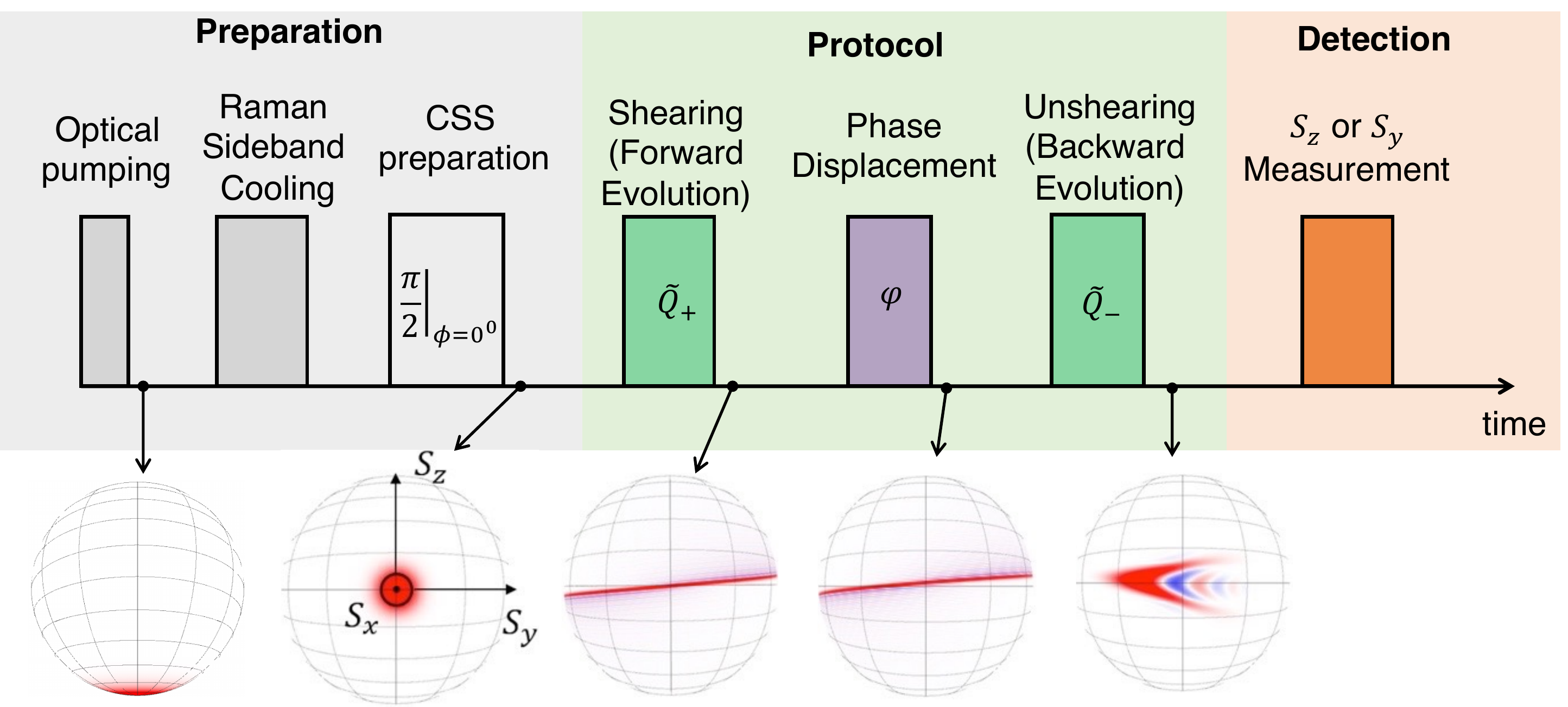}
\caption{
\textbf{Experimental sequence.} The three stages of the experiment are represented in different color-shaded areas in the upper part: preparation, protocol, and detection. Bloch spheres (bottom) show the collective atomic state after the indicated process has been performed in time. The time axis is not to scale. 
}
\label{fig:figSM_1}
\end{figure}

After performing Raman sideband cooling that leaves the atoms spin polarized in the state $\ket{\uparrow}{=}\ket{m_{I}{=}+1/2}$, the ensemble is prepared into a Coherent Spin State (CSS) of the two magnetic sublevels of the ground state $\ket{^1S_0}$ ($\ket{m_I{=}{\pm}1/2}$). We drive this transition by using radiofrequency (RF) pulses generated by a pair of coils in the presence of an external magnetic field $B_z{=}13.6$~G, corresponding to a Larmor frequency of $2\pi{\times}10.2$~kHz.
The Rabi frequency of the RF pulses is $2\pi \times 208(2)$~Hz.
After the CSS is prepared, the SATIN protocol starts. Fig.~\ref{fig:figSM_1} shows the three experimental stages of our protocol.

\paragraph*{State measurement}
The value of $S_z$ is obtained from the difference $S_z{=}(N_\uparrow{-}N_\downarrow){/}2$ of the populations $N_\uparrow$ and $N_\downarrow$ of the states $\ket{\uparrow}$ and $\ket{\downarrow}$, respectively.
We first measure $N_\uparrow$ through the vacuum Rabi splitting of the cavity mode $2g{\approx}\sqrt{N_\uparrow\eta\kappa\Gamma}$ when the empty cavity is resonant with the transition $\ket{\uparrow}\rightarrow\ket{e}=\ket{^3P_1, m_F=+3/2}$~\cite{braverman2019near}. 
Here $\kappa{=}2\pi{\times}530(10)~$kHz is the cavity linewidth and $\Gamma{=}2\pi{\times}184(1)~$kHz the linewidth of the atomic transition. The Rabi splitting is measured by scanning the laser frequency and detecting the cavity transmission as a function of the frequency. 

The resolution of a single measurement, normalized to the SQL, is given by $\sigma_d^2= \frac{1}{S_0}\mathrm{Var}(S_{z1}-S_{z2})$ where $S_{z1}$ and $S_{z2}$ are two state measurements performed after a single CSS preparation. We obtain $\sigma_d^2{=}0.15\pm0.02$ and it remains constant within the whole range of atom numbers used in this experiment. 
Since all atoms have the same coupling to the cavity, the atom number $N$ inferred from the Rabi splitting equals the real number of atoms in the cavity. 

\paragraph*{Single-atom cooperativity measurement}

We can calculate the single-atom cooperativity $\eta$ accurately from our cavity parameters \cite{Kawasaki2019, braverman2019near, Pedrozo2020entanglement}. With a measured finesse of $F=11400$ and atoms loaded $0.246\pm0.004~$mm from the micromirror \cite{Kawasaki2019}, we expect a single-atom cooperativity $\eta=7.8\pm0.2$.

We also verify it by measuring the spin projection noise via the cavity as a function of the collective cooperativity $N\eta$.
For a coherent spin state (CSS) prepared at the equator of the generalized Bloch sphere, the measured variance of the difference $\eta S_z = \frac{\eta N_\uparrow-\eta N_\downarrow}{2}$ is 
\begin{equation}
\mathrm{var}(\eta S_z)=(N\eta)\frac{\eta (1+\sigma_d^2)}{4},
\end{equation}
where we have also included the contribution due to the measurement noise $\sigma_d^2$.
The latter contribution is obtained through the variance of the difference between two measurements after a single CSS preparation.

\begin{figure}[hbtp]
\setlength{\unitlength}{1\textwidth}
\includegraphics[width=100mm,scale=.9]{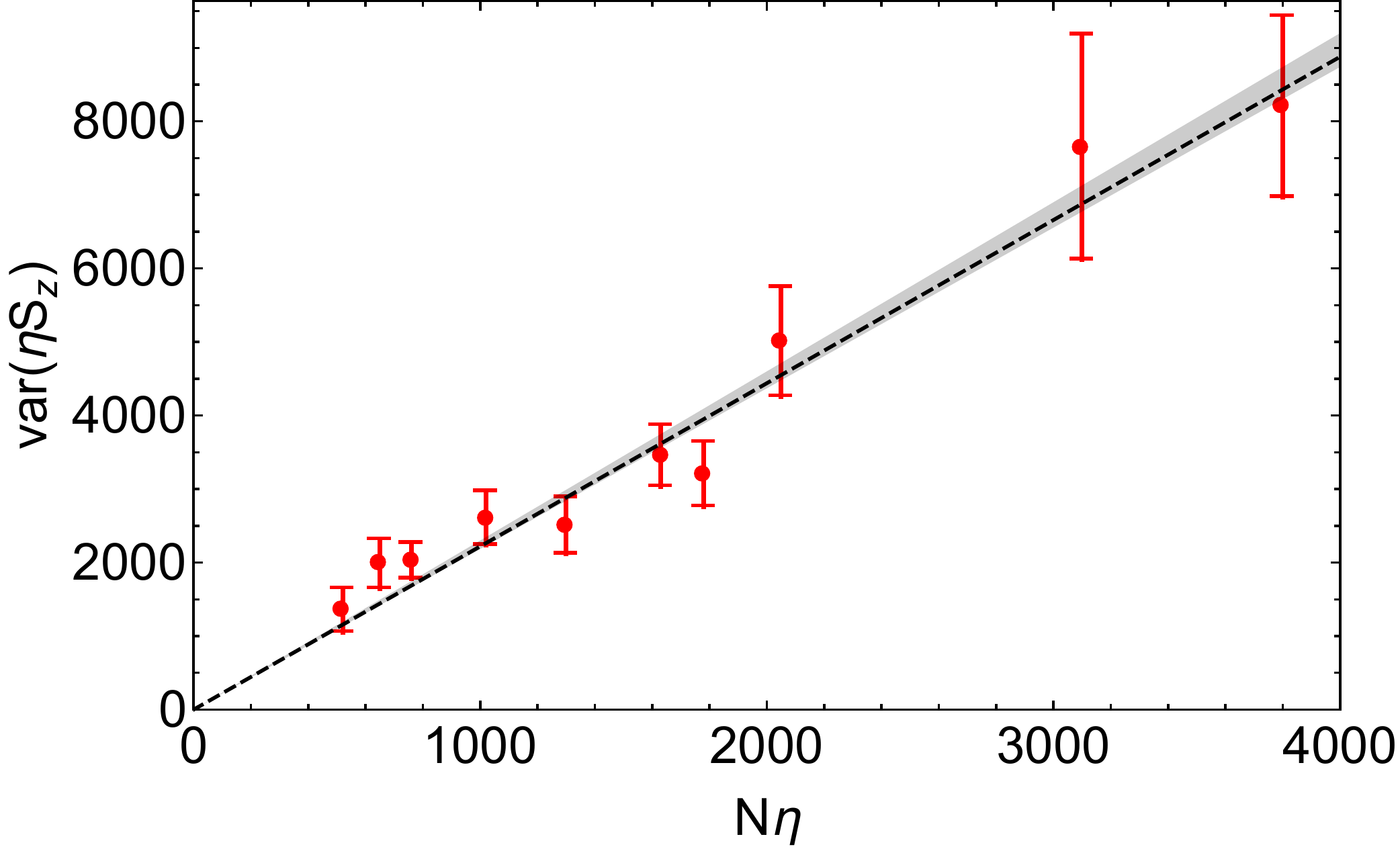}
\caption{
\textbf{Single-atom cooperativity.} The linearity of the data indicates the absence of classical sources of noise, which would manifest as a quadratic dependence of the measured variance on collective cooperativity $N \eta$. The dashed line represents the linear regression fit with slope $\eta(1+\sigma_d^2)/4 = 2.2\pm0.1$. The gray band denotes the expected projection noise given by the single-atom cooperativity calculated from our cavity parameters $\eta=7.8\pm0.2$. Each data  point corresponds to the mean value obtain from 50 to 150 experimental realizations. The error bars correspond to 1 $\sigma$ and are the standard error of a Gaussian distribution: $\sigma_{s}^2/(n-1)$, where $\sigma_{s}$ is the sample variance and $n$ is the number of experimental realizations.
}
\label{fig:figSI_0}
\end{figure}

Plotting the variance of $\eta S_z$ as a function of the collective cooperativity results, in the absence of classical sources of noise, in a linear graph with slope $\eta(1+\sigma_d^2)/4$ (see Fig.~\ref{fig:figSI_0}).
Since the measurement resolution normalized to the CSS spin projection noise $\sigma_d^2=0.15\pm0.02$ is a constant, we obtain the single-atom cooperativity $\eta = 7.7\pm0.3$ by fitting the data to a linear model, in good agreement with our direct calculation from the measured cavity parameters. When a quadratic fitting term is included to account for possible technical noise, we recover the same cooperativity $\eta=7.7 \pm 0.9$ within error bars.

The single-atom cooperativity inferred form spin projection noise measurements agrees with the calculated one within error bars (see Fig.~\ref{fig:figSI_0}).

\begin{figure}[hbtp]
\setlength{\unitlength}{1\textwidth}
\includegraphics[width=100mm,scale=1]{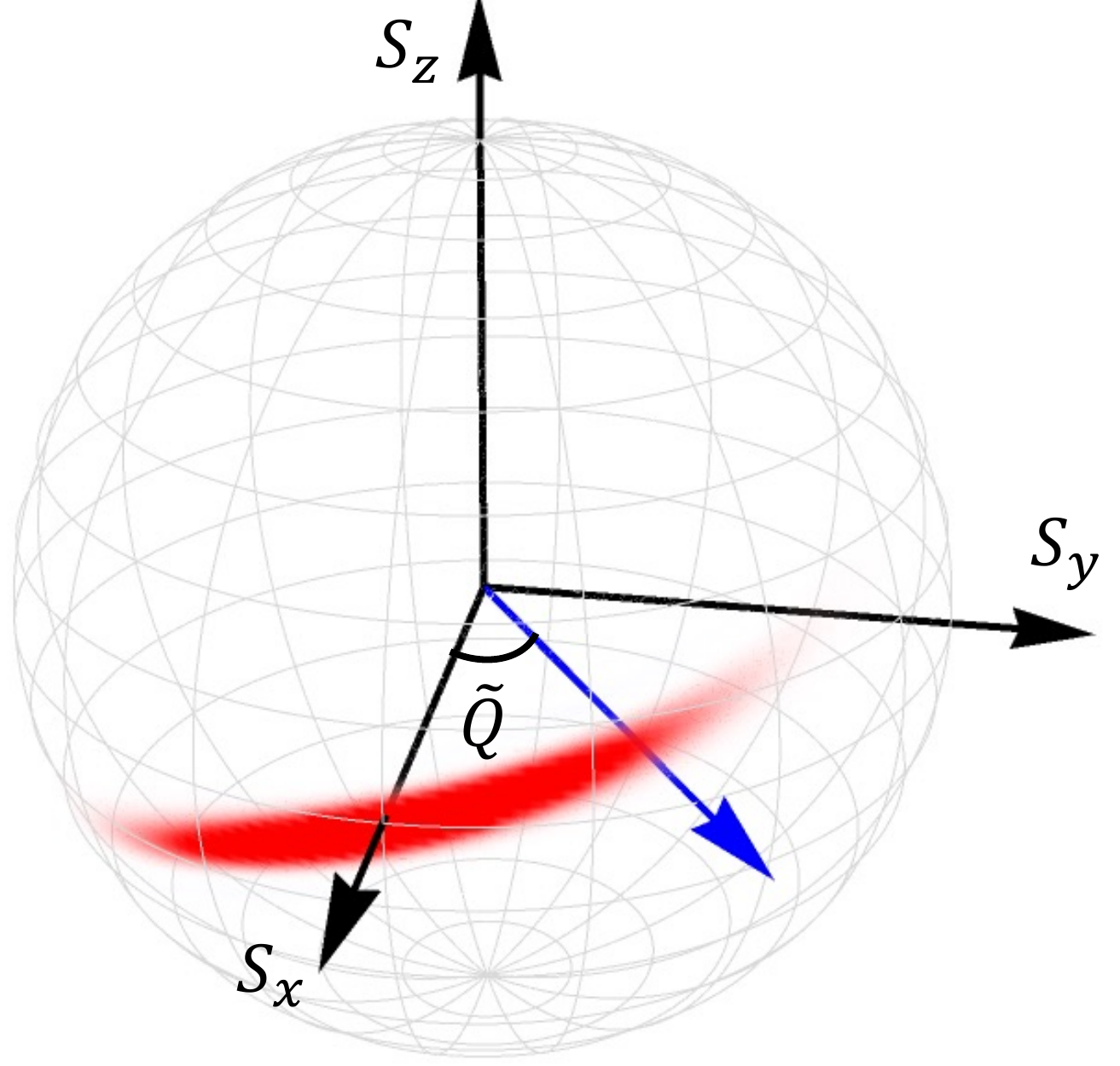}
\caption{
\textbf{Squeezed spin distribution on the generalized Bloch sphere.} The normalized shearing strength $\tilde{Q}$ represents the angle subtended by the sheared distribution with respect to the $x$-axis, along which the initial CSS was prepared.
}
\label{fig:figSM_09}
\end{figure}

\paragraph*{Cavity-induced one-axis twisting}
The squeezing Hamiltonian is the result of the interaction of the atomic ensemble with the single-mode light inside an optical cavity, which is given by (Eq. 15 in \cite{li2021collective}):
\def\y{x_a}
\begin{equation}\label{eq:SqueezingHamiltonian}
\begin{split}
    \hat{H}_{\textrm{dip}}&=-\left(\hat{S}_z+S\right)\eta\frac{|\hat{\mathcal{E}}_c|^2}{\omega}\frac{\pi}{\mathcal{F}}\mathcal{L}_d(\y) \\
    &= - \hbar \Omega \hat{n}_c \left(\hat{S}_z+S\right), 
\end{split}
\end{equation}
where $\hat{\mathcal{E}}_c$ is the amplitude of the intracavity field, $\mathcal{F}$ is the cavity Finesse, $\mathcal{L}_d(x_a) = -\frac{x_a}{1+x_a^2}$ is the dispersive Lorentzian profile  with $x_a\equiv2\Delta/\Gamma$ the normalized detuning of the probe laser from the atomic resonance ($\Delta=\omega_l-\omega_a$) with respect to the natural linewidth of the transition ($\Gamma$). In this expression  $\Omega = \pi\eta \mathcal{L}_d(\y)\kappa /\mathcal{F}$ represents the light shift per photon inside the cavity with $\kappa$ the cavity linewidth, and $\hat{n}_c=|\hat{\mathcal{E}}_c|^2/(2\hbar\omega\kappa)$ the photon number inside the cavity.  
The second term of this Hamiltonian represents a global rotation that is canceled by the spin echo sequence. This leads to a Hamiltonian $\hat{H} = - \hbar \Omega \hat{n}_c \hat{S}_z$ that depends on $S_z$ and the number of photons inside the cavity.\\
When the CSS is close to the equator of the Bloch sphere, $\langle \hat{S}_z\rangle=0$, we can expand the photon number $\hat{n}_c(\hat{S}_z)$ in terms of $S_z$, and write the Hamiltonian as:
\begin{equation}\label{eq:CavityAtomsStarkShiftTwoLevelAtomTaylorSeries}
\hat{H} = -\hbar \Omega \hat{S}_z \sum_{j=0}^\infty \frac{S_z^j}{j!}\left(\frac{\partial^j \hat{n}_c}{\partial S_z^j}\right)_{S_z=0}.
\end{equation}
The zero-order term of the expansion ($\hat{H}_0= -\hbar \Omega\langle{\hat{n}_c}\rangle \hat{S}_z$) represents a rotation of the collective state around the $z$-axis of the Bloch sphere, which is also canceled by the spin echo sequence. The first-order term of the expansion,
\begin{equation}\label{eq:TwistingHamiltonian}
\hat{H}_1 = -\hbar \chi \hat{S}_z^2,
\end{equation}
is the known one-axis twisting Hamiltonian \cite{Kitagawa1993}. This term represents the effective spin-spin interaction mediated by light and produces a rotation of the atomic spin around the $z$-axis that is proportional to $S_z$, producing the squeezed distribution of the collective atomic state, as shown in Fig.~\ref{fig:figSM_09}.
%
The twisting or shearing parameter is given by
\begin{equation}\label{eq:CavityAtomsSz2}
\chi = - \eta \mathcal{L}_d(x_a) \left(1-x_a+ \frac{N}{2}\eta\right)\mathcal{T}_0 \frac{\mathcal{S} }{N/2},
\end{equation}
which is proportional to the scattered photon number into free space $\mathcal{S}$. Here, $\mathcal{T}_0$ is the power transmitted through a symmetric and lossless cavity, and is given:
\begin{equation}\label{eq:SymmetricTransmission}
    \mathcal{T}_0=\frac{\left|\mathcal{E}_t\right|^2}{\left|\mathcal{E}_r\right|^2}=\frac{1}{(1+\frac{N}{2}\eta\mathcal{L}_a(x_a))^2+(x_c+\frac{N}{2}\eta\mathcal{L}_d(x_a))^2},
\end{equation}
and we have defined the dispersive and absorptive Lorentzian profiles $\mathcal{L}_d(x) = -\frac{x}{1+x^2}$, and $\mathcal{L}_a(x) = \frac{1}{1+x^2}$, respectively. $x_c\equiv 2\delta/\kappa$ is the detuning of the probe beam from the cavity resonance frequency normalized to the cavity linewidth.

%
To implement the effective cavity-induced OAT Hamiltonian, Eq. 1, we first tune the frequency of the high-finesse cavity $\omega_c$ in resonance with the $\ket{\uparrow} \rightarrow \ket{e} = \ket{^3P_1, m_F=+3/2}$ transition at frequency $\omega_a$, so that strong coupling of the cavity field to the atoms results in vacuum Rabi splitting (Fig. 1). 
A pulse of light with frequency $\omega_l$ tuned to the slope of a Rabi peak (Fig. 1) will pass through the cavity and interact with the atoms, resulting in the first-order phase shift $\beta S_z$ and shearing $\chi S_z^2$~\cite{braverman2019near, li2021collective}.
After cancelling the first-order phase shift with a spin echo sequence~\cite{Pedrozo2020entanglement}, the system evolution can be described by the OAT Hamiltonian $\hat{H} = \chi S_z^2$.

\begin{figure}[hbtp]
\setlength{\unitlength}{1\textwidth}
\includegraphics[width=100mm,scale=1]{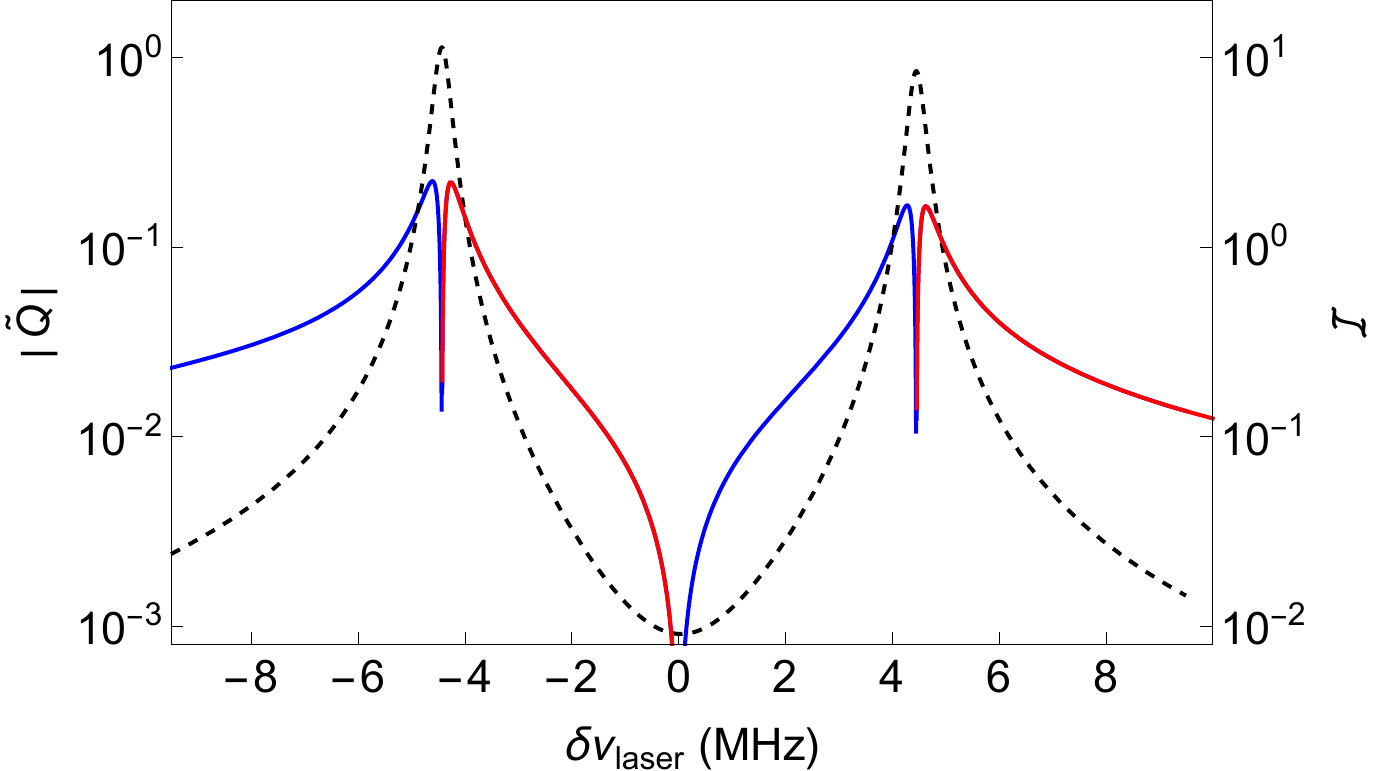}
\caption{
\textbf{Relevant parameters for our squeezing protocol.} Excess broadening $\mathcal{I}$ (black dashed line) per scattered photon and shearing strength $|\tilde{Q}|$ (solid line). The red and blue parts of the solid line represents positive and negative values of $\tilde{Q}$, respectively, which lead to to forward and backward evolutions in time. In this figure, we have used the experimental parameters $N{=}220$ and $\eta{=}7.7$. For illustration purposes, contrast loss has not been included.
}
\label{fig:figSM_08}
\end{figure}

It is useful to express the action of the OAT Hamiltonian in terms of the normalized twisting parameter \begin{equation}
\tilde{Q}\equiv\sqrt{N}\chi\tau \label{eq:Qtilde}
\end{equation}
$\tilde{Q}$ is the rms angle subtended by the state on the Bloch sphere, and $\tau$ is the entangling time, i.e., the action time of the OAT Hamiltonian.
Using~\cite{li2021collective}, the twisting parameter is expressed as

\begin{equation}\label{eq:3levelQ}
    \tilde{Q} =\frac{n_{tr}^{tot}}{\sqrt{N}}\mathcal{L}_d(x_a)\mathcal{L}_a(x_a)\frac{\frac{N}{2}\eta^2(1+\frac{N}{2}\eta-x_a x_c)}{\big(1+\frac{N}{2}\eta\mathcal{L}_a(x_a)\big)^2+\big(x_c+\frac{N}{2}\eta\mathcal{L}_d(x_a)\big)^2}
\end{equation}
where $n_{tr}^\mathrm{tot}$ is the total number of photons transmitted through the cavity.
We notice that $\tilde{Q}(-x_a,-x_c)=-\tilde{Q}(x_a, x_c)$. This means the sign of the shearing parameter (i.e., the "shearing direction") can be switched by changing the sign of the detuning of the laser frequency from the atomic (and cavity mode) transition frequency $\omega_a=\omega_c$. Hence, from Eq.~\ref{eq:Qtilde}, $\chi=\tilde{Q}/(\tau\sqrt{N})$, it follows that the sign of the Hamiltonian is also switched, as represented in Fig.~\ref{fig:figSM_08}.

Similarly, we evaluate the additional light-induced broadening $ \mathcal{I}$ of the phase noise of the atomic state. ($ \mathcal{I}=1$ means that the additional broadening equals the CSS variance.)
\begin{equation}\label{eq:3levelF}
 \mathcal{I} = 2\,n_{tr}^\mathrm{tot}\mathcal{L}_a^2(x_a)\frac{(N/2) \eta^2(1+N \eta/2 +x_a^2)}{\big(1+(N/2)\eta\mathcal{L}_a(x_a)\big)^2+\big(x_c+(N/2)\eta\mathcal{L}_d(x_a)\big)^2}.
 \end{equation}

We also consider the effects on $\tilde{Q}$ and $\mathcal{I}$ of atoms populating the $\ket{\downarrow}$ level~(see~\cite{li2021collective}~for details).
Atoms in the $\ket{\downarrow}$ state have a similar contribution to the polarizability as atoms in $\ket{\uparrow}$, but the corresponding transition will be detuned due to the Zeeman shift $\Delta_Z\approx20~$MHz between the excited sublevels $\ket{^3P_1, m_F=+1/2}$ and $\ket{e}=\ket{^3P_1, m_F=+3/2}$, due to the $14$~G magnetic field applied along the $z$-axis (see Fig.~\ref{fig:Scheme}).
%

\paragraph*{Quantum Noise in $S_y$ quadrature}

We first consider the phase of the spin vector $\tau_y$ in the absence of contrast loss of the signal, defined as $\tau_y\equiv\sqrt{2|\langle S\rangle|}\mathrm{arcsin}\left(S_y/|\langle S\rangle|\right)$ for $\tau_y<\pi/2$.

After letting an initial CSS evolve forward and backward under the OAT Hamiltonian, the variance of $\Delta\tau_y^2$ is
is
\begin{equation}
   \Delta\tau_y^2 = \frac{1+\mathcal{I_\mathrm{tot}}}{2 S_0}+ \tilde{Q}_\mathrm{tot}^2,
   \label{eq:OATphase_noise}
\end{equation}
where $S_0$ is the generalized Bloch sphere radius, $\tilde{Q}_\mathrm{tot}=\tilde{Q}_{+}+\tilde{Q}_{-}$, and  $\mathcal{I}_\mathrm{tot}=\mathcal{I}_{+}+\mathcal{I}_{-}$, i.e., the sum of the excess broadenings induced by the twisting ($\mathcal{I}_+$) and untwisting ($\mathcal{I}_-$) procedures.
This results in a spin variance normalized to the CSS of
\begin{equation}
    \frac{2 \mathrm{var}(S_y)}{S_0}\equiv\sigma^2_y=\frac{1}{2}+\frac{1}{2}\,\exp\left(-2 \Delta\tau_y^2\right).
\end{equation}

\paragraph*{Contrast loss}
During the twisting and untwisting processes, there is contrast reduction, or equivalently, a shrinking of the radius of the Bloch sphere associated with the collective atomic spin. The reduction of the length of the spin vector due to photon scattering is given by
\begin{equation}
    \mathcal{C}_{sc}\equiv\frac{|\langle S\rangle|}{S_0}=\exp\left(-2\frac{n_{sc}(\tilde{Q}_{+},\tilde{Q}_{-})}{N}\right),
\end{equation}
where $n_{sc}(\tilde{Q}_{+},\tilde{Q}_{-})$ is the total number of photons scattered into free space to generate both $\tilde{Q}_{+}$ and $\tilde{Q}_{-}$.
%
\begin{figure}[hbtp]
\setlength{\unitlength}{1\textwidth}
\includegraphics[width=89mm,scale=.9]{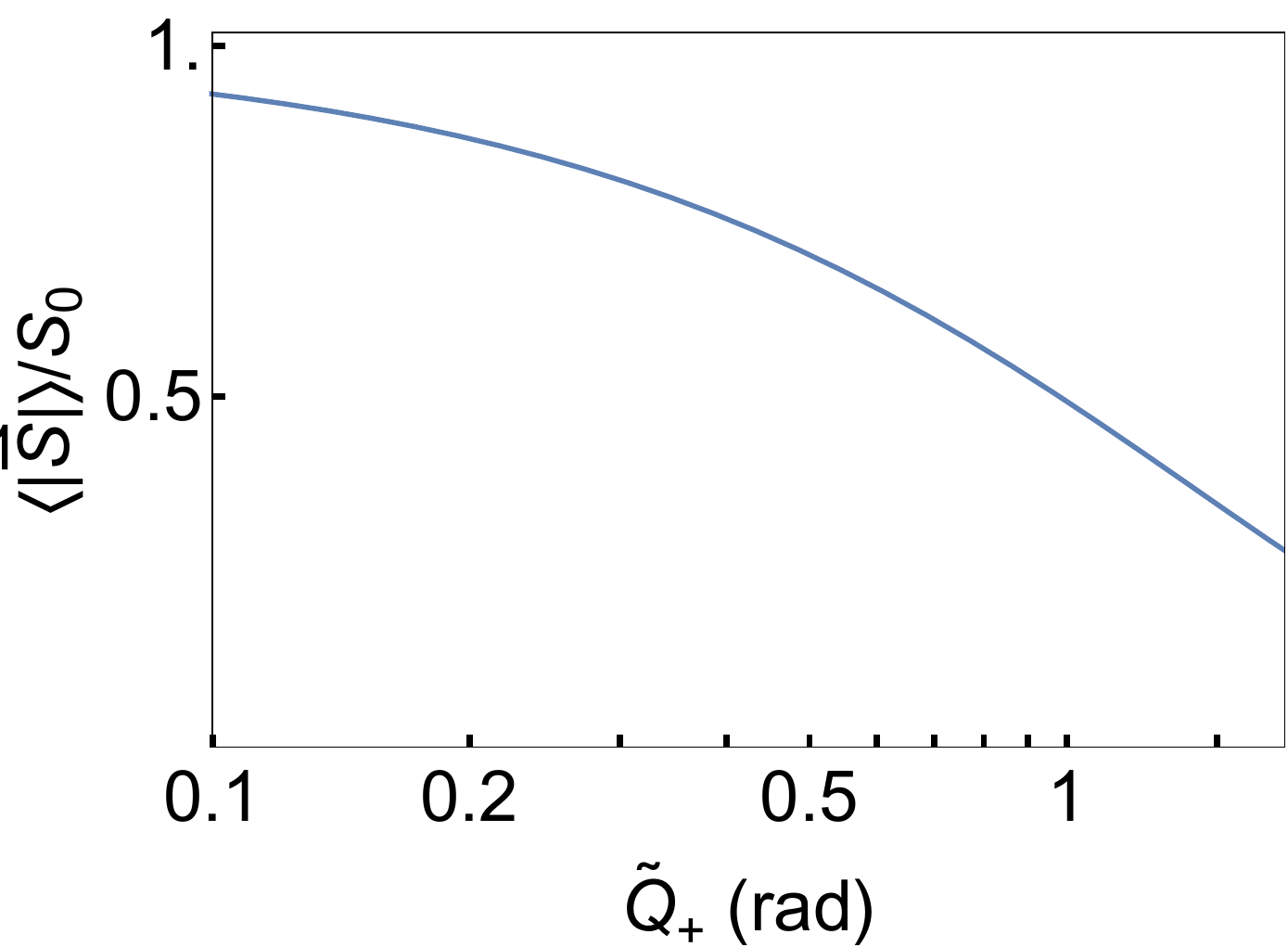}
\caption{
\textbf{SATIN contrast loss.} Contrast reduction in an optimized SATIN protocol as a function of the twisting strength $\tilde{Q}_{+}$. "Optimized SATIN" means that the entangling light detuning is chosen in order to maximize the protocol's metrological gain. It is worth noting that, under this optimization condition, the contrast reduction is independent of the atom number. 
}
\label{fig:figSI_1}
\end{figure}

To evaluate the spin noise projection on the $S_y$-quadrature, we first consider the spin phase noise of the coherent sub-ensemble of atoms, i.e. of the atoms that have not scattered a photon into free space. 
The spin-vector length of this sub-ensemble is $|\langle S\rangle|$, and the resulting spin phase variance is  
\begin{equation}
   \Delta\tau_y^2 = \frac{1+\mathcal{I}_\mathrm{tot}}{2 C_{sc}S_0}+ \tilde{Q}_\mathrm{tot}^2.
   \label{eq:OATphase_noise_cp}
\end{equation}
It is worth noting that both $\tilde{Q}_{\mathrm{tot}}$ and $\mathcal{I}_{\mathrm{tot}}$ are not affected by the contrast loss; they solely depend on the $S_z$ projection which here we can consider as remaining unchanged by the entangling light.

\begin{figure}[hbtp]
\setlength{\unitlength}{1\textwidth}
\includegraphics[width=89mm,scale=.9]{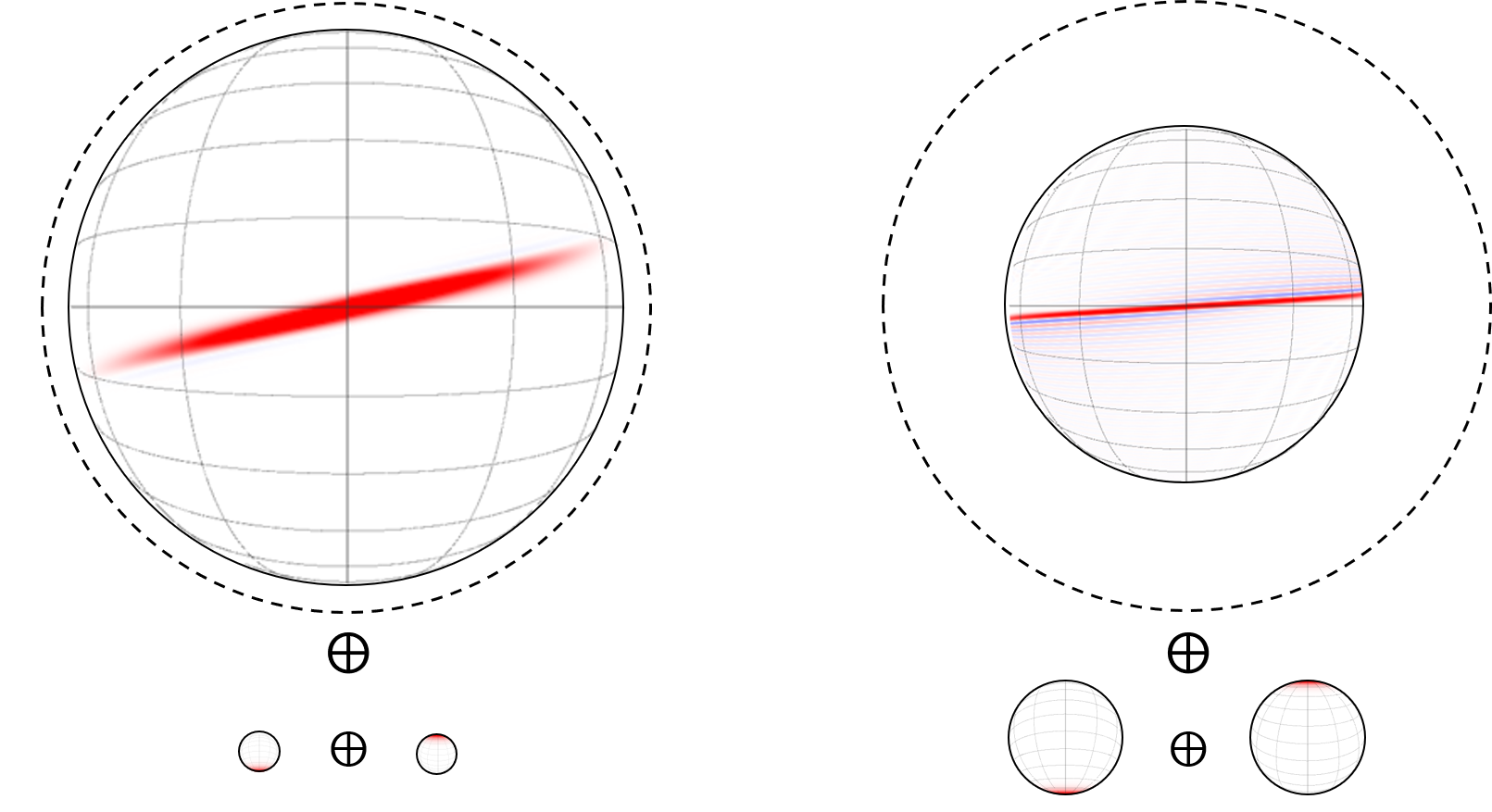}
\caption{
\textbf{Graphical representation of the model used to describe contrast loss due to scattering of photons into free space.} The atomic spin can be decomposed into the coherent signal (large Bloch sphere) and the signals of the sub-ensembles of atoms that due to photon scattering have been projected into the spin states $\ket{\uparrow}$ or $\ket{\downarrow}$, and that have lost any coherence.
The \textbf{left} figure corresponds to $\tilde{Q}_+=0.3$ (mostly Gaussian distribution) while the \textbf{right} figure is calculated for $\tilde{Q}_+=1.3$.}
\label{fig:figSI_2}
\end{figure}

Considering the spin variance contribution of the atoms that have scattered a photon and whose states are uncorrelated with the ensemble, we obtain the normalized spin variance
\begin{equation}
    \frac{\mathrm{var}(S_y)}{S_0/2}=1 - \mathcal{C}_{sc} + S_0 \mathcal{C}_{sc}^2\left\{1-\exp\left[-2\,\left(\frac{1+\mathcal{I}_\mathrm{tot}}{2\mathcal{C}_{sc}S_0}+ \tilde{Q}_\mathrm{tot}^2\right)\right]\right\} 
\end{equation}

Under the Holstein-Primakoff approximation ($\mathcal{I} \ll N$, $\tilde{Q} \ll 1$), the variance of the state reduces to
\begin{equation}
    \frac{\mathrm{var}(S_y)}{S_0/2} = 1 + 2 S_0 \mathcal{C}_{sc}^2\,\tilde{Q}_\mathrm{tot}^2 +\mathcal{C}_{sc}\mathcal{I}_\mathrm{tot}.
\end{equation}

\paragraph*{Signal Amplification}


The expression for the signal amplification as a function of the twisting strength is derived in~\cite{Davis2016}, and for $N \gg 1$ reads
\begin{equation}
    m(\tilde{Q}) \approx \mathcal{C}_{sc}(\tilde{Q}) \cdot N \cdot \sin \left( \frac{\tilde{Q}}{\sqrt{N}} \right) \cdot \cos ^{N} \left( \frac{\tilde{Q}}{\sqrt{N}} \right), 
\end{equation}
where $\tilde{Q}=\tilde{Q}_{+}=\tilde{Q}_{-}$.
Note that for small $\tilde{Q} \ll 1$, the maximal signal enhancement is obtained when the state is displaced not along $S_y$, but at an angle $\theta \approx \arctan(1/m)$ to it, or for an optimized $\tilde{Q}_{-}>\tilde{Q}_{+}$~\cite{schulte2020ramsey}.
However, here we are interested in $\tilde{Q} \approx 1$, where $\theta \approx 0$, and we induce or measure displacements directly along $S_y$.

\paragraph*{Light-shift during OAT. }
During the OAT process, the zero-order term of the Hamiltonian (5), induces an absolute light-shift of $\phi_\mathrm{lightshift} {\approx}8\,\pi \times \tilde{Q}$.
This contribution is canceled by a spin echo sequence~\cite{braverman2019near}.

The light-shift is induced by the average number of photons $n_\mathrm{avg}$ transmitted through the cavity. Under optimized detuning, for every atom number $N$, the average photon number is given by ${n_\mathrm{avg}\approx 1.6 \times N \times \tilde{Q}}$.

\paragraph*{Computation of Wigner quasiprobability distribution functions}
We compute Wigner quasiprobability distributions on the Bloch sphere following efficient computation methods presented in Ref.~\cite{koczor2020fast}.

\end{document}